\documentclass[pra,twocolumn,longbibliography,preprintnumbers,superscriptaddress]{revtex4-2}
\usepackage[dvipsnames]{xcolor}
\definecolor{mediumpersianblue}{rgb}{0.0, 0.4, 0.65}
\definecolor{persianred}{rgb}{0.8, 0.2, 0.2}
\usepackage[colorlinks=true,citecolor=mediumpersianblue,linkcolor=persianred, urlcolor=persianred]{hyperref}
\usepackage{physics}
\usepackage{graphicx}
\usepackage{amsmath}
\usepackage{amssymb}
\usepackage{bm}
\usepackage{xspace,slashed}
\usepackage{enumitem}
\usepackage{algorithm2e}
\SetKwComment{Comment}{/* }{ */}

\usepackage{multirow}
\usepackage[utf8]{inputenc}
\usepackage{bm}
\usepackage{braket}
\usepackage{array}
\usepackage{tikz}
\usetikzlibrary{quantikz2}

\usepackage{xspace,slashed}
\usepackage{array}
\usepackage{fixme}

\usepackage{csquotes}
\usepackage[caption=false]{subfig}
\usepackage{utfsym}

\newcommand{\Qibolab}{\texttt{Qibolab}\xspace}
\newcommand{\Qibo}{\texttt{Qibo}\xspace}
\newcommand{\Qibocal}{\texttt{Qibocal}\xspace}

\fxsetup{
    status=draft,
    layout=inline,
    theme=color
}
\definecolor{fxnote}{rgb}{1.000,0.0000,0.0000}

\begin{document}

\title{Real-time error mitigation for variational optimization on quantum hardware}

\preprint{TIF-UNIMI-2023-10, CERN-TH-2023-207}

\newcommand{\MIaff}{TIF Lab, Dipartimento di Fisica, Universit\`a degli Studi di
  Milano and INFN Sezione di Milano, Milan, Italy.}

\newcommand{\TII}{Quantum Research Centre, Technology Innovation Institute, Abu Dhabi, UAE.}

\newcommand{\CERNaff}{CERN, Theoretical Physics Department, CH-1211
  Geneva 23, Switzerland.}

\newcommand{\ANU}{School of Computing, The Australian National University, Canberra, ACT, Australia}

\newcommand{\IFT}{Instituto de F\'isica Te\'orica, UAM-CSIC, Universidad Aut\'onoma de Madrid, Cantoblanco, Madrid, Spain}

\author{Matteo Robbiati}
\affiliation{\MIaff}
\affiliation{\CERNaff}
\author{Alejandro Sopena}
\affiliation{\IFT}
\author{Andrea Papaluca}
\affiliation{\ANU}
\affiliation{\TII}
\author{Stefano Carrazza}
\affiliation{\MIaff}
\affiliation{\CERNaff}
\affiliation{\TII}

\begin{abstract}
  In this work we put forward the inclusion of error mitigation routines in the process of training
  Variational Quantum Circuit (VQC) models. In detail, we define a Real Time Quantum Error Mitigation (RTQEM)
  algorithm to assist in fitting functions on quantum chips with VQCs.
  While state-of-the-art QEM methods cannot address the exponential loss concentration induced by noise in current devices, we demonstrate
  that our RTQEM routine can enhance VQCs' trainability by reducing the corruption of the loss function. We tested the algorithm by simulating and deploying the fit of a
  monodimensional {\it u}-Quark Parton Distribution Function (PDF) on a superconducting single-qubit device, and we further analyzed the scalability of
  the proposed technique by simulating a multidimensional fit with up to 8 qubits.
\end{abstract}

\maketitle
\label{sec:introduction}

In the era of Noisy Intermediate Scale Quantum (NISQ)~\cite{Preskill_2018, Bharti_2022}
devices, Variational Quantum Algorithms (VQA) are the Quantum Machine Learning (QML) 
models that appear more promising in the near future. They have several 
concrete applications already validated,
such as electronic structure modelization in quantum chemistry~\cite{Peruzzo_2014, 
McClean_2016, Bauer_2016, Jones_2019}, for instance.
Different VQA ansätze have been proposed, such as the QAOA~\cite{Farhi_2019}, but they all share as
foundation a Variational Quantum Circuit (VQC) consisting of several parameterized gates whose
parameters are updated during training.

Hardware errors and large execution times corrupt the landscape in various ways, 
such as changing the position of the minimum or the optimal value of the loss function, 
hindering NISQ~\cite{Preskill_2018, Bharti_2022} devices' applicability in practice 
for certain algorithms. Furthermore, VQC models are known to suffer from the presence of 
Noise-Induced Barren Plateaus (NIBPs)~\cite{Wang_2021} in the optimization space, 
leading to vanishing gradients. NIBPs are fundamentally different from 
the noise-free barren plateaus discussed in 
Refs~\cite{mcclean_barren_2018, cerezo_cost_2021, arrasmith_effect_2021, holmes_connecting_2022, ragone_unified_2023, diaz_showcasing_2023}. 
In fact, approaches designed to tackle noise-free barren plateaus do 
not seem to effectively address the issues posed by NIBPs~\cite{Wang_2021}.

To overcome these limitations we either have to build fault-tolerant
architectures carrying a usable amount of logical qubits, or exploit the available 
NISQ hardware by mitigating its results from the noise.
While the first solution might require significant technical advances,
the second one is often achieved with the help of quantum error mitigation (QEM)~\cite{kandala_error_2019}.
Exponential loss concentration cannot be resolved with error 
mitigation~\cite{wang2021error}, but it is possible to improve trainability by 
attempting to reduce the loss corruption.
Therefore, we define here an algorithm to perform Real-Time Quantum Error 
Mitigation (RTQEM) alongside a VQA-based QML training process. 

In this work, we use the Importance Clifford Sampling (ISC) method~\cite{Qin_2023}, a learning-based quantum error mitigation 
procedure~\cite{PRXQuantum.2.040330}. The core business of the learning-based QEM techniques
is to approximate the noise with a parametric map which, once learned, can be used 
to clean the noisy results~\cite{urbanek_mitigating_2021,a_rahman_self-mitigating_2022,farrell_preparations_2023,ciavarella_quantum_2023,farrell_scalable_2023}. Linear maps have the
potential to improve overall trainability by addressing challenges imposed by loss corruptions while not affecting 
loss concentration itself~\cite{wang2021error}. The map's parameters are learned during the QML training every time 
the noise changes above a certain arbitrarily set threshold.

\begin{figure*}[ht]
  \centering
    \includegraphics[width=1\linewidth]{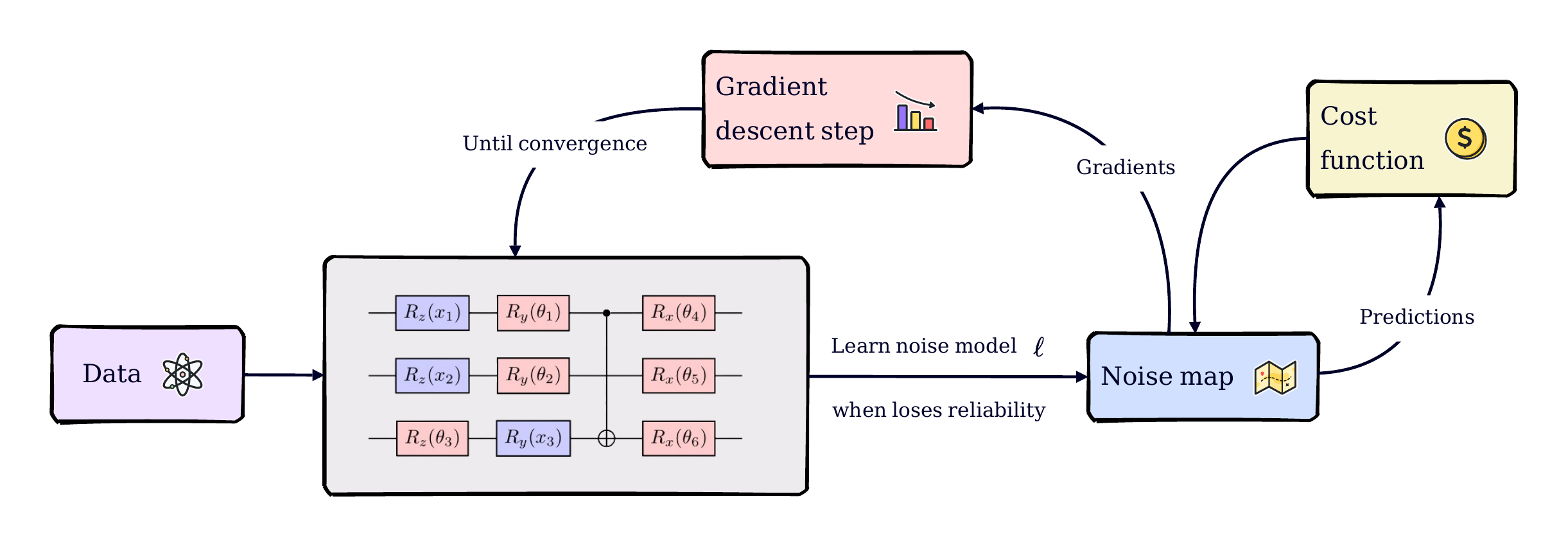}
    \caption{The RTQEM pipeline involves training a variational quantum circuit on a noisy quantum device using a gradient descent method enhanced with error mitigation. 
    Specifically, the ICS algorithm is used to learn a noise map to mitigate both the gradients and the final predictions. 
    If the noise changes above a certain threshold, the noise map is re-learned.}
    \label{fig:algo_scheme}
  \end{figure*}

We apply the RTQEM strategy to a series of mono-dimensional and multi-dimensional
regression problems.
Firstly, we train a VQC to tackle a particularly 
interesting High Energy Physics (HEP) problem: 
determining the Parton Distribution Function (PDF) of the $u$-quark, one of the proton
contents. In a second step, we define a multi-dimensional target to study the impact 
of the RTQEM procedure when the VQA involves an increasing number of qubits.

Data re-uploading~\cite{P_rez_Salinas_2020} is used to encode data into the model, 
while we implement a hardware-compatible Adam~\cite{kingma2017adam} optimizer for the training. 
We calculate gradients with respect to the variational parameters using the 
Parameter Shift Rule~\cite{Mitarai_2018, Schuld_2019} (PSR). This optimization scheme 
is ideal for studying the performance of RTQEM, as the PSR formulas require 
a number of circuits to be executed which scales linearly with the number of parameters.
The greater the number of executions, the better our algorithm must be to enable training of the model.

This setup is then used to perform the full $u$-quark PDF fit on two different 
superconducting quantum devices hosted in the Quantum 
Research Center (QRC) of the Technology Innovation Institute (TII).  

The whole work has been realized using the Qibo framework, which offers \Qibo~\cite{Efthymiou_2021, 
Efthymiou_2022, Carrazza_2023, stavros_efthymiou_2023_7736837} as high-level language
API to write quantum computing algorithms, \Qibolab~\cite{efthymiou2023qibolab, 
stavros_efthymiou_2023_7748527, carobene2023qibosoq} as quantum control tool and 
\Qibocal~\cite{pasquale2023opensource, andrea_pasquale_2023_7662185, pedicillo2023benchmarking} to perform 
quantum characterization and calibration routines.

The outline is as follows. In Section~\ref{sec:methodology} we summarize the process of quantum
computing with the VQC paradigm, providing also details about the ansatz and the PSR rule we make use
of to train the model. In Section~\ref{sec:noise}, we discuss the impact of noise on the training process and 
provide an overview of the error mitigation strategy we employed to counteract these effects. Finally, we report the results of our
experiments both with noisy simulations and real superconducting qubits deployment in Section~\ref{sec:validation}.

\section{\label{sec:methodology}Methodology}

\subsection{A snapshot of Quantum Machine Learning}

In the following we are going to consider Supervised Machine Learning problems for
simplicity, but what presented here can be easily extended to other Machine Learning (ML) paradigms
in the quantum computation context. Quantum Machine Learning (QML) arises when using Quantum Computing (QC) tools to
tackle ML problems~\cite{Biamonte_2017, Schuld_2014, Mitarai_2018}. 

In the classical scenario, given an $n$-dimensional input variable $\bm{x}$, a 
parametric model is requested to 
estimate a target variable $y$, which is related to $\bm{x}=(x_1,\ldots,x_n)$ through some hidden law
$y=g(\bm{x})$. The model estimations $y_{\rm est}$ are then compared with some measured 
ground truth data $y_{\rm meas}$ by evaluating a loss function $J(y_{\rm est}, y_{\rm meas})$,
which quantifies the capability of the model to provide an estimate of the underlying law $g$.
 We consider the output variable $y$ as mono-dimensional for simplicity,
but in general it can be multi-dimensional.
The variational parameters $\bm{\theta}$ of the model are then optimized to minimize (or maximize)
the loss function $J$, leading, in turn, to better predictions $y_{\rm est}$. 

In Quantum Machine Learning, we translate this paradigm to the language of quantum computing. 
In particular, parametric quantum gates, such as rotations,
are used to build Variational Quantum Circuits (VQC)~\cite{chen2020variational},
which can be used as parametric models in the machine learning process.  
Once a parametric circuit $\mathcal{U}(\theta)$ is defined, it can be applied to a 
prepared initial state $\ket{\psi_0}$ of a quantum system to obtain the final 
state $\ket{\psi_f}$, which is used to evaluate the expected value of an 
arbitrary chosen observable $\mathcal{O}$,
\begin{equation}
f(\bm{\theta})_{\mathcal{O}} = \braket{\psi_0 |\,\mathcal{U}^{\dagger}(\bm{\theta}) \mathcal{O} 
\, \mathcal{U}(\bm{\theta})| \psi_0} \ .
\label{eq:qml_estimator}
\end{equation}
Various methods exist to embed input data into a QML process~\cite{lloyd2020quantum, Havl_ek_2019, incudini2022structure}; 
in this work, we employ the re-uploading strategy~\cite{P_rez_Salinas_2020}.
The estimates of $y$ can be obtained by calculating expected values of the form~\eqref{eq:qml_estimator}.
Finally, the circuit's parameters are optimized to minimize (or maximize) a 
loss function $J$,
pushing $f$ as close as possible to the unknown law $g$.

\subsection{A variational circuit with data-reuploading}

The data-reuploading~\cite{P_rez_Salinas_2020} method is built by defining a parameterized layer made of fundamental uploading gates
which accepts the input data $\bm{x}$ to be uploaded. Then, the re-uploading of 
the variable is achieved by building a circuit composed of a sequence of uploading layers. 
\begin{figure*} 
  \begin{tikzcd}[row sep = 2]
  \lstick{$\ket{0}$} & \gate{L(x_1| \bm{\theta}_{1,1})} & \ctrl{1} & \qw       & \qw       & \qw      & \targ{}   &   \qw & \ \cdots \ & \gate{L(x_1| \bm{\theta}_{l,1})} & \ctrl{1} & \qw       & \qw       & \qw      & \targ{} & \meter{} \\
  \lstick{$\ket{0}$} & \gate{L(x_2| \bm{\theta}_{1,2})} & \targ{}  & \ctrl{1}  & \qw       & \qw      & \qw       &   \qw & \ \cdots \ & \gate{L(x_2| \bm{\theta}_{l,2})} & \targ{}  & \ctrl{1}  & \qw       & \qw      & \qw & \meter{}  \\
  \lstick{$\ket{0}$} & \gate{L(x_3| \bm{\theta}_{1,3})} & \qw      & \targ{}   & \ctrl{1}  & \qw      & \qw       &   \qw & \ \cdots \ & \gate{L(x_3| \bm{\theta}_{l,3})} & \qw      & \targ{}   & \ctrl{1}  & \qw      & \qw & \meter{} \\
  \lstick{$\ket{0}$} & \gate{L(x_4| \bm{\theta}_{1,4})} & \qw      & \qw       & \targ{}   & \ctrl{1} & \qw       &   \qw & \ \cdots \ & \gate{L(x_4| \bm{\theta}_{l,4})} & \qw      & \qw       & \targ{}   & \ctrl{1} & \qw & \meter{} \\
  \lstick{$\ket{0}$} & \gate{L(x_n| \bm{\theta}_{1,n})} & \qw      & \qw       & \qw       & \targ{}  & \ctrl{-4} &   \qw & \ \cdots \ & \gate{L(x_n| \bm{\theta}_{l,n})} & \qw      & \qw       & \qw       & \targ{}  & \ctrl{-4} & \meter{} \\ 
  \end{tikzcd}
\caption{Circuit ansatz to reupload the data $\bm{x}$ with $n$ qubits and $l$ layers.}
\label{fig:qpdf}
\end{figure*}
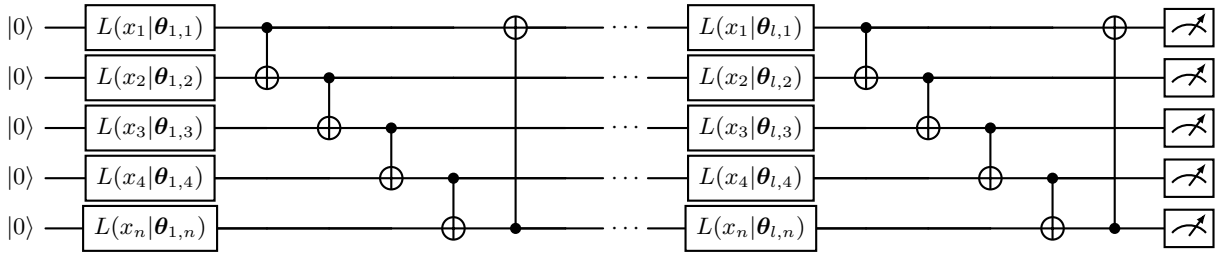
Inspired by~\cite{P_rez_Salinas_2021}, we build our ansatz by defining the following fundamental uploading gate,
\begin{equation}
L(x_j| \bm{\theta}_{l,j}) = R_z(\theta_3\,x_j + \theta_4) R_y(\theta_1\, \kappa(x_j) + \theta_2) \ ,
\label{eq:uploading_layer}
\end{equation}
where $x_j$ is the $j$-th component of the input data and with $\bm{\theta}_{l,j}$
we denote the four-parameters vector composing the gate which uploads $x_j$ at the ansatz layer
$l$. The information $x_j$ is uploaded twice in each $L$, first in the $R_z$ and 
second in the $R_y$  through an activation function $\kappa(x_j)$.
To embed the $n$ components of $\bm{x}$ into the ansatz, we build a $n$-qubit circuit $\mathcal{U}$ based on the Hardware Efficient Ansatz, 
where the single-qubit gates are the fundamental uploading gates, and entanglement is generated with CNOT gates,
as shown in Fig.~\ref{fig:qpdf}. 
We calculate $f$~\eqref{eq:qml_estimator} as the expected value of the Pauli observable $\sigma_z^{\otimes n}$ on the final state $\mathcal{U}\left(\ket{0}^{\otimes n}\right)$.

\subsection{Gradient descent on hardware}

Gradient-based optimizers~\cite{Rumelhart1986LearningRB, kingma2017adam, JMLR:v12:duchi11a, ruder2017overview} 
are commonly employed in machine learning problems, 
particularly when using Neural Networks~\cite{Schmidhuber_2015} (NNs) as models. 
In the QML context, VQCs are utilized to construct Quantum Neural Networks~\cite{Abbas_2021}, 
which serve as quantum analogs of classical NNs. 
Consequently, we are naturally led to believe that methods based on gradient calculation could be effective.

\subsubsection{The parameter shift rule} 

In order to perform a gradient descent on a NISQ device we need a 
method that is robust to noise and executable on hardware. This cannot be done as in the classical 
case following a back-propagation~\cite{Rumelhart1986LearningRB} of the information 
through the network. We would need to know the $f$ values during the propagation, 
but accessing this information would collapse the quantum state. 
Moreover, standard finite-differences methods are not applicable due to the shot noise when executing the circuit a finite number of times.
An alternative method is the so called Parameter Shift 
Rule~\cite{crooks2019gradients, Schuld_2019, Wierichs_2022, Mari_2021, Banchi_2021} (PSR), which
enables the evaluation of quantum gradients directly on the hardware~\cite{Schuld_2019}.
Given $f$ as introduced in~\eqref{eq:qml_estimator} and considering a single parameter 
$\mu \in \bm{\theta}$ affecting a single gate whose hermitian generator 
has at most two eigenvalues, the PSR allows for
the calculation
\begin{equation}
\partial_\mu f = r \bigl( f(\mu^+) - f(\mu^-)\bigr) \ ,
\label{eq:psr}
\end{equation}
where $\pm r$ are the generator eigenvalues, $\mu^{\pm}=\mu \pm s$ and $s=-\pi/4r$.
In other words, the derivative is calculated by executing twice the same circuit 
$\mathcal{U}(\bm{\theta})$ in which the parameter $\mu$ is shifted backward and forward of $s$.
A remarkable case of the PSR involves rotation gates, 
for which we have $r=1/2$ and $s=\pi/2$~\cite{Mitarai_2018}. 

\subsubsection{Evaluating gradients of a re-uploading model}

In order to perform a gradient-based optimization, we first need to calculate the 
gradient of a loss function $J$ with respect to the variational parameters of
the model. Then, the derivatives are used to perform an optimization step
in the parameters' space by following the steepest direction of the gradient,
\begin{equation}
\bm{\theta}_{t+1} = \bm{\theta}_t - \eta \nabla J(\bm{\theta}_i) \ ,
\label{eq:sgd_step}
\end{equation} 
where $\eta$ is the learning rate of the gradient descent algorithm.
Since our QML model is a circuit in which the variational parameters
are rotation angles, such derivatives can be estimated by the PSR~\eqref{eq:psr}.
However, even in the simplest case, 
this kind of procedure can be computationally expensive, since for
each parameter we need two evaluations of $f$, as
illustrated in~\eqref{eq:psr}. Given a VQC with $p$ variational parameters, a training set size of $N_{\rm data}$, 
and a budget of $N_{\rm shots}$ for each function evaluation, the total computational cost amounts to
$2pN_{\rm shots}N_{\rm data}$ circuit executions.
This high number of evaluation is useful for testing the effectiveness
of error mitigation routines, which can be applied to every function evaluation of the algorithm. We followed the same optimization strategy described
in~\cite{Sweke_2020, robbiati2022quantum}, defining a Mean-Squared Error loss
function,
\begin{equation}
  J_{\rm mse}(\bm{x}^i|\bm{\theta}) = \frac{1}{N_{\rm data}}\sum_i^{N_{\rm data}} 
  \bigl[ f(\bm{x}^i,\bm{\theta}) - g(\bm{x}^i) \bigr]^2 \ ,
  \label{eq:loss}  
\end{equation}
where the superscript denotes the $i$-th variable $\bm{x}$ of the dataset. 
Note that this differs from the subscripts used so far to denote the components of the variable $\bm{x}$.

Our total execution time is dominated by the effect of circuit switching and network 
latency costs rather than shot cost. Therefore, we prefer to reduce the number 
of iterations at the expense of increasing the number of shots per iteration. 
In this context, the Adam~\cite{kingma2017adam} optimizer stands out due to its 
robustness when dealing with complex parameters landscapes.

\section{\label{sec:noise}Noise on quantum hardware}

Recognizing the impact of noise on the optimization landscape is crucial in practical quantum computing implementations. 
In the presence of a general class of local noise models, for many important ansätzes 
such as Hardware Efficient Ansatz (HEA), the gradient decreases exponentially with the depth of the circuit $d$. 
Typically, $d$ scales polynomially with the number of qubits $n$, causing the gradient to decrease exponentially in $n$. 
This phenomenon is referred to as a Noise-Induced Barren Plateau (NIBP)~\cite{wang_noise-induced_2021}. 
NIBPs can be seen as a consequence of the loss function converging around the value associated with the maximally mixed state. 
Furthermore, noise can corrupt the loss landscape in various ways such as changing the position of the minimum.

In order to quantify these effects, we consider a noise model composed of local Pauli channels acting on qubit $j$ before and after each layer of our ansatz,
\begin{equation}
  \mathcal{P}_j(\sigma) = q_j\sigma \,
\end{equation}
where $-1<q_x,q_y,q_z<1$ and $\sigma$ denotes the local Pauli operators $\{\sigma_x,\sigma_y,\sigma_z\}$. The overall channel is $\mathcal{P} = \bigotimes_j^N \mathcal{P}_j$. 
We also include symmetric readout noise $\mathcal{M}$ made of single-qubit bit-flip channels with bit-flip probability $(1-q_M)/2$. This results in the noisy expectation value,
\begin{equation}
  f_{\text{noisy}} = \operatorname{Tr}\bigl[\sigma_z^{\otimes n}\left(\mathcal{M} \circ \mathcal{P} 
  \circ L_l \circ \cdots \circ L_1 \circ 
  \mathcal{P}\right)(\ket{\psi_0} \bra{\psi_0}) \bigr] \ .
  \label{eq:cost_bound}
\end{equation}
The NIBP translates into a concentration of the expectation value around $0$~\cite{wang_noise-induced_2021},
\begin{equation}
  \abs{f_{\text{noisy}}} < 2q_M^nq^{2l + 2}\left(1-\frac{1}{2^n}\right) \ .
\end{equation}
Certain loss functions exhibit noise resilience, {\it i.e.} their minimum remains in the same position under the influence of certain noise models, even though its value may change. 
Contrarily, our loss function~\eqref{eq:loss} is not noise resistant. We aim to explore the extent to which it is possible to mitigate the noise and enhance the training process of VQCs with non-resistant loss functions.

\subsection{\label{sec:error_mit}Error Mitigation}

Recent research~\cite{li_efficient_2017, temme_error_2017, lowe_unified_2021, Czarnik_2021, sopena_simulating_2021, van_den_berg_model-free_2022} 
has focused on developing methods to define unbiased estimators 
of the ideal expected values leveraging the knowledge about the noise that we 
can extract from the hardware. However, these estimators are also affected by 
exponential loss concentration, implying that NIBPs cannot be resolved without 
requiring exponential resources through error mitigation~\cite{wang2021error}.

In the regime where loss concentration is not severe, it is also not straightforward 
for error mitigation to improve the resolvability of the noisy loss landscape, 
thus alleviating exponential concentration.

The variance of the error-mitigated estimators is typically higher than that of 
the mean estimator~\cite{cai_quantum_2022}, setting up a trade-off between the 
improvement due to bias reduction and the worsening caused by increased variance.
While most methods have a negative impact on resolvability, linear ansatz methods~\cite{urbanek_mitigating_2021,Qin_2023,a_rahman_self-mitigating_2022,farrell_preparations_2023,ciavarella_quantum_2023,farrell_scalable_2023} show potential due to their neutral impact under global depolarizing noise~\cite{wang2021error}. 
Among all of them, Importance Clifford Sampling~\cite{Qin_2023} stands out for its ability to handle single qubit dependent noise, its scalability, and its resource cost.

\subsubsection{\label{sec:ics}Importance Clifford Sampling (ICS)}

Suppose we want to estimate the expected value of an observable $\mathcal{O}$ 
for the state $\rho$ prepared by a quantum circuit $\mathcal{C}^0$. 
In a realistic situation we are going to obtain a noisy expected value $\langle 
\mathcal{O}\rangle_{\rm noisy}^0$ different from the true one $\langle \mathcal{O}\rangle^0$. 
The idea behind Importance Clifford Sampling (ICS) is to generate a set of $m$ training Clifford
circuits $\mathcal{S}=\big\{\mathcal{C}^i\big\}_{i=1}^m$ with the same circuit frame as
the original one $\mathcal{C}^0$. The classical computation of noiseless expected values of Clifford circuits 
is efficient~\cite{aaronson_improved_2004, pashayan_fast_2022}. This enables us to compute the ideal 
expected values $\big\{\langle\mathcal{O}\rangle^i\big\}_{i=1}^m$ through simulation, 
as well as the noisy expected values $\big\{\langle\mathcal{O}\rangle^i_{\text{noisy}}\big\}_{i=1}^m$ when evaluating them on hardware.

\begin{figure}
  \centering
    \includegraphics[width=1\linewidth]{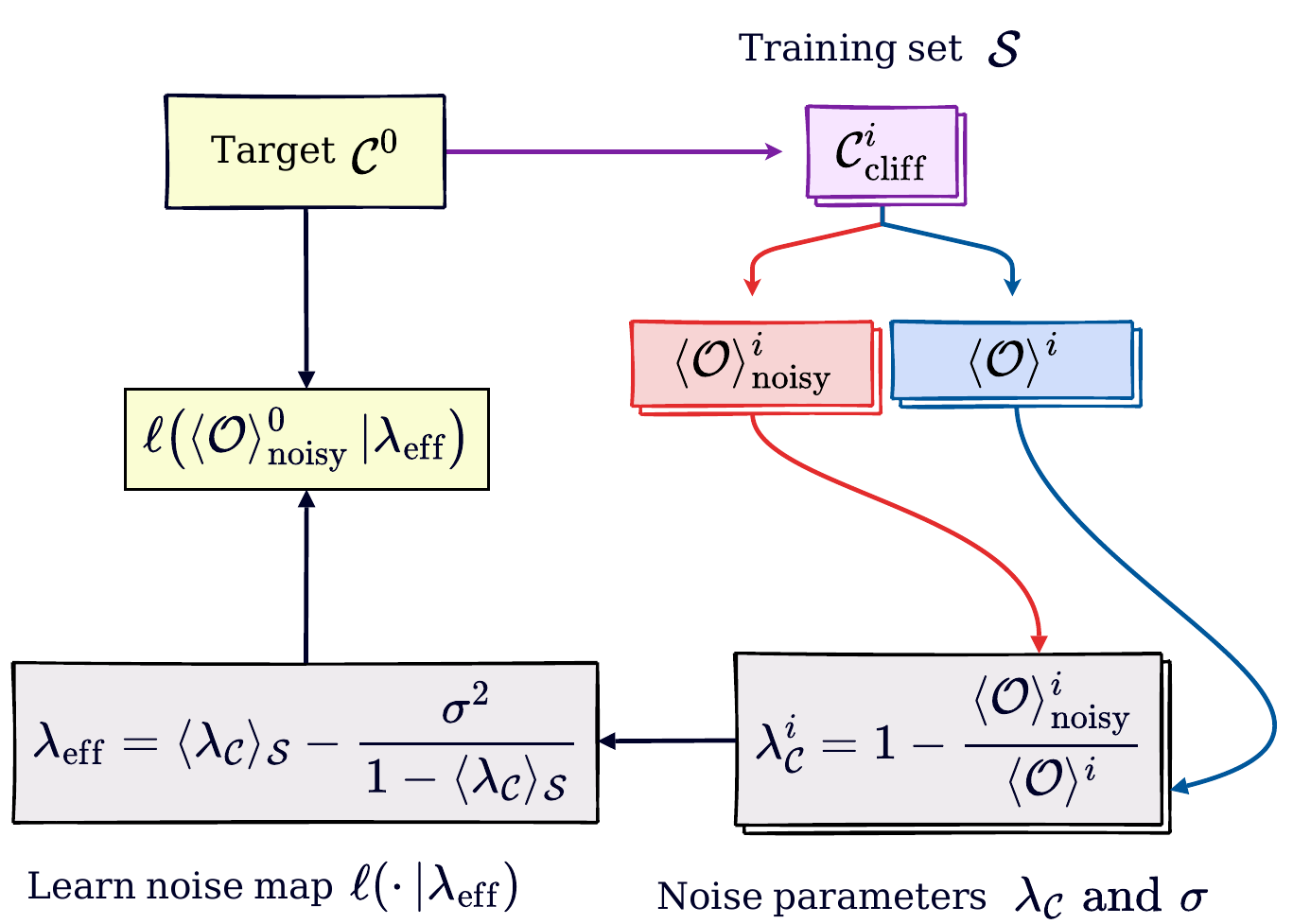}
    \caption{Importance Clifford Sampling is a learning based error mitigation algorithm that uses a set of Clifford circuits to learn a noise map to mitigate the expected value of a given observable.}
    \label{fig:ics}
\end{figure}

When $\mathcal{O}$ is a Pauli string, the noise-free expected values will concentrate on $-1$, $0$, $1$~\cite{aaronson_improved_2004}.
Furthermore, as discussed in~\cite{Qin_2023}, not all the Clifford circuits are
error sensitive. In particular, we only need circuits whose expected values on Pauli's 
are $\pm 1$. We refer to these circuits as \textit{non-zero} circuits for simplicity.
Unfortunately, sampling \textit{non-zero} circuits is exponentially rare when the 
number of qubits increases, thus a strategy has to be defined to efficiently build
a suitable training set. We follow the ICS algorithm~\cite{Qin_2023}, 
in which \textit{non-zero} circuits are built by adding a layer of Pauli gates to \textit{zero} circuits. 
These gates can be merged with the ones following so that the depth does not increase.

The generated set is then used to train a model to learn a mapping between $\langle \mathcal{O}\rangle_{\text{noisy}}$ and $\langle \mathcal{O}\rangle$. 
The structure of the model $\ell$ can be inspired by considering the action of a global depolarizing channel with depolarizing parameter~$\lambda$,
\begin{equation}
  \langle \mathcal{O}\rangle_{\text{noisy}} = (1-\lambda)\langle 
    \mathcal{O}\rangle + \frac{\lambda}{d}\operatorname{Tr}(\mathcal{O}) \ ,
\end{equation}
where $d=2^n$ denotes the dimension of the Hilbert space and $0<\lambda<4^n/(4^n-1)$. 
Focusing on Pauli strings and allowing $\lambda$ to take any value, we arrive at the phenomenological error model,
\begin{equation}
  \langle \mathcal{O}\rangle_{\text{noisy}}^i = (1-\lambda_\mathcal{C}^i)\langle 
    \mathcal{O}\rangle^i \ ,
\end{equation}
from which we can calculate $\lambda_\mathcal{C}^i$ for each circuit in the training set. 
This set of depolarizing parameters, characterized by the mean value $\lambda_0=\langle \lambda_\mathcal{C}\rangle_\mathcal{S}$ and standard deviation $\sigma$, allows to define an effective depolarizing parameter for mitigating the initial circuit,
\begin{equation}
  \lambda_{\text{eff}} = \lambda_0 - \frac{\sigma^2}{1-\lambda_0} \ .
\label{eq:lambda_eff}
\end{equation}
This translates into the noise map,
\begin{equation}
  \ell \bigl( \langle O \rangle | \lambda_{\text{eff}} \bigr) = \frac{(1-\lambda_0)}{(1-\lambda_0)^2 + \sigma^2} \langle O \rangle_{\rm noisy} \ . 
\label{eq:noise_map} 
\end{equation}
The average depolarizing rate $\lambda_0$ scales proportionally with the number of gates, while the standard deviation $\sigma$ is proportional to its square root~\cite{Qin_2023}. 
This implies that the model performs better as the circuit depth increases.

The noise map~\eqref{eq:noise_map} effectively handles symmetric readout noise, but fails with asymmetric noise. 
For these situations, we employ Bayesian Iterative Unfolding (BIU)~\cite{nachman_unfolding_2020} to mitigate measurement errors in advance.

A schematic representation of the described algorithm is reported in Fig.~\ref{fig:ics}.

\section{\label{sec:rtqem}The RTQEM algorithm}

We implement an Adam optimization mitigating both gradients and predictions following 
the procedure presented in Sec.~\ref{sec:ics}.

In a real quantum computer, small fluctuations of the conditions over time, such as temperature, may result in a change of the shape of the noise sufficient to deteriorate results. Therefore, we compute a 
metric
\begin{equation}
D(z, \ell(z)) = |z - \ell(z)|
\label{eq:metric_noise}
\end{equation}
at each optimization iteration, which quantifies the distance
between a target noiseless expected value $z$ and the mitigated estimation $\ell(z)$.
These expected values are calculated over a single \textit{non-zero} test circuit to maximize the bias.
If an arbitrary set threshold value $\varepsilon_{\ell}$ is exceeded, the noise map is relearned from scratch.  
A schematic representation of the proposed procedure is reported in Alg.~\ref{alg:rtqem}.

\RestyleAlgo{ruled} 
\begin{algorithm}\label{alg:rtqem}
\vspace{0.1cm}
\vspace{0.1cm}
\textbf{Set} $N_{\rm update}, N_{\rm epoch}, k=0$ \;
\textbf{Initialize} VQC parameters $\bm{\theta}_k$, noise map $\ell$ \;
\textbf{Define} target noiseless expectation value $z$ \;
\textbf{Define} metric $D(z, \ell(z))$ to check $\ell$ reliability\;

\vspace{0.1cm}
\vspace{0.1cm}

\For{$k < N_{\rm epochs}$}{
  \If{$D(z, \ell(z))>\varepsilon_{\ell}$}{
    learn $\ell_k$\;
    $\ell \leftarrow \ell_k$\;
  }
  compute $\ell({\bm{y}}_{\rm est})$\;
  calculate $J\bigl[\ell(\bm{y}_{\rm est}), \bm{y}_{\rm meas}\bigr]$\;
  \For{$p \in \bm{\theta_k}$}{
    compute $\ell(\partial_{p}J)$ via PSR\;
  }
  $\bm{\theta}_{k+1}$ $\leftarrow$ Adam$(\bm{\theta}_{k})$;
}
\caption{RTQEM}
\end{algorithm}

\section{\label{sec:validation}Validation}

We propose two different experiments to test the RTQEM algorithm 
introduced above. Firstly, in Sec.~\ref{sec:simulations}, we simulate the training 
of a VQC on both a single and a multi-qubit noisy device. Whereas, in Sec.~\ref{sec:hardware}, 
the same procedure is deployed on a superconducting single-qubit chip. 
The programs to reproduce such simulations can be found at~\cite{rtqem}.

\subsection{\label{sec:simulations}Simulation}

In this section, we benchmark different levels of error mitigation by conducting both noisy 
and noiseless classical simulations with $N_{\rm shots}=10000$ shots as outlined in Tab.~\ref{tab:simulations}.
The VQC shown in Fig.~\ref{fig:qpdf} is used as ansatz and the noise is described 
by the noise model presented in Section~\ref{sec:noise}. We first consider a static-noise 
scenario in Section~\ref{sec:static_noise}, while in Section~\ref{sec:evolving_noise}
we let the noise vary over time.

\begin{table}[h]
\centering
\begin{tabular}{cccc}
\hline \hline 
\textbf{Training} & \textbf{Noise} & \textbf{RTQEM} & \textbf{QEM at the end} \\
\hline
Noiseless & \usym{2613}   & \usym{2613}   & \usym{2613} \\
Noisy     & \usym{1F5F8}  & \usym{2613}   & \usym{2613} \\
fQEM     & \usym{1F5F8}  & \usym{2613}   & \usym{1F5F8} \\
RTQEM     & \usym{1F5F8}  & \usym{1F5F8}  & \usym{1F5F8} \\
\hline \hline
\end{tabular}
\caption{\label{tab:simulations}Summary of the tested simulation configurations.}
\end{table}

\subsubsection{\label{sec:static_noise}Static-noise scenario}

The following simulations are performed using a static local Pauli noise model 
where we set the following noise parameters  $q_x=0.007$, $q_y=0.003$, $q_z=0.002$ and $q_M=0.005$. 

\begin{figure}
  \centering
    \includegraphics[width=0.95\linewidth]{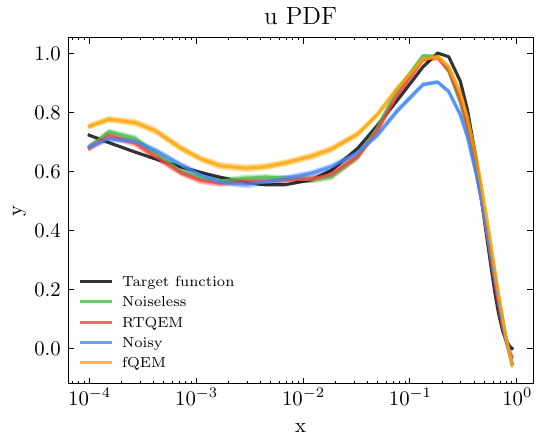}
    \includegraphics[width=0.96\linewidth]{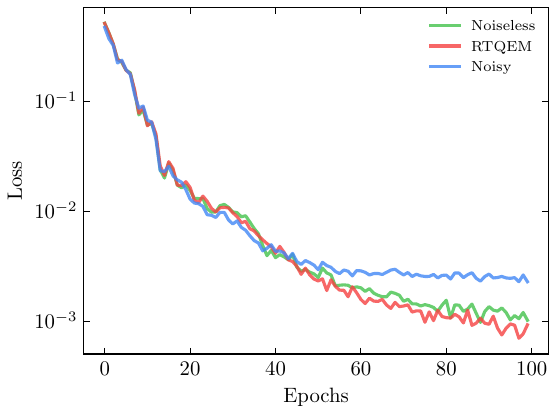}
    \caption{Estimates of $u$-quark PDF associated to $N_{\rm data}=50$ momentum fraction
    values sampled logarithmically in $[0, 1]$. The NNPDF4.0 measures (black line)
    are compared with results obtained through noiseless simulation (green line), 
    noisy simulation (blue line), noisy simulation with mitigation applied to the 
    final predictions (yellow line) and real-time mitigated noisy simulation (red line).
    The effective depolarizing parameter $\lambda_{\text{eff}}$ is $0.09\pm0.01$.
    The final predictions are computed averaging on $N_{\rm runs}=100$ predictions 
    calculated for each of the $N_{\rm data}$ points. The confidence intervals 
    are obtained using one standard deviation from the mean. The bottom plot shows 
    the loss function history for each training scenario.}
    \label{fig:qpdf_simulation}
  \end{figure}

We first consider a one-dimensional target, namely, the $u$-quark 
Parton Distribution Function (PDF) for a fixed energy scale $Q_0$ with varying
momentum fraction $x$ sampled from the interval $[0,1]$. 
A logarithmic sampling is used to improve the resolution of the $x\sim(0,0.1)$ range
where the shape of the function is more rugged.
The corresponding PDF values are provided by the NNPDF4.0 grid~\cite{Ball_2022}.
We address this first target by constructing a four-layer single-qubit circuit, following the ansatz depicted in Fig.~\ref{fig:qpdf}. 
The results, shown in Fig.~\ref{fig:qpdf_simulation}, illustrate that the RTQEM approach enables the training to converge to the correct solution.

To avoid the u-quark PDF comfortably resting below the bound~\eqref{eq:cost_bound} and thereby 
disguising the effect of the NIBP, we opted to expand it to cover the range $[0, 1]$. 
One might wonder whether a similar, but opposite, trick could be employed in case that the bound
intercepts the target function. Therefore, compressing the function to make it lie
below the bound and avoid any sort of limitations on the predictions. While this is a
perfectly viable method in theory, it is essentially pointless in practice. The compression
of the function, indeed, will also increase the precision needed to resolve it, which
translates in a larger number of shots required by each prediction~\cite{wang2021error}. 

The noisy simulation is clearly limited by loss concentration~\eqref{eq:cost_bound}, 
which caps the predictions at around $y\simeq0.85$.
This limit is also noticeable in Fig.~\ref{fig:qpdf_simulation}, where the loss value cannot decrease below the threshold with unmitigated training. 
Attempting to correct the predictions post-training (f-QEM) allows access to the region above the bound, but does not enhance the fit.
This is expected, as the noise shifts the position of the minima of the loss function 
making it impossible to retrieve the true minimum with a final rescaling pass alone.
However, by gradually cleaning up the loss function landscape during the training, 
the correct minimum is recovered, and the fit converges to the target function.

 \begin{figure*}[t]
  \centering
          \includegraphics[width=0.28\linewidth]{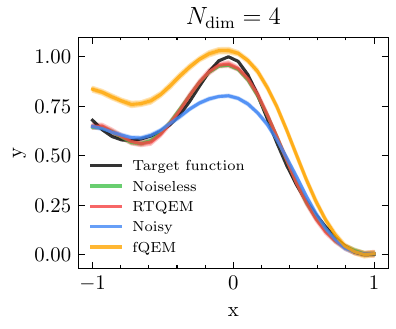}%
          \includegraphics[width=0.28\linewidth]{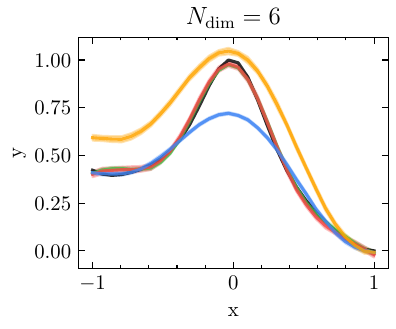}%
          \includegraphics[width=0.28\linewidth]{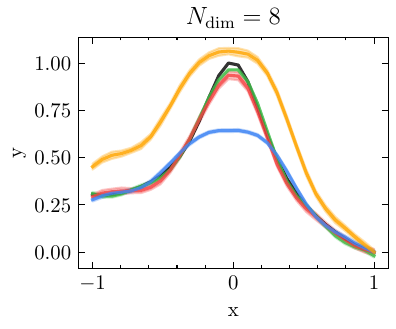}
          \includegraphics[width=0.28\linewidth]{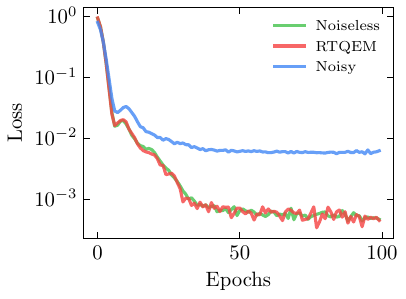}%
          \includegraphics[width=0.28\linewidth]{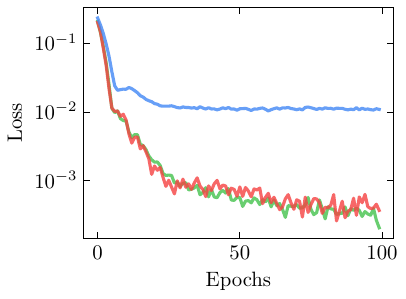}%
          \includegraphics[width=0.28\linewidth]{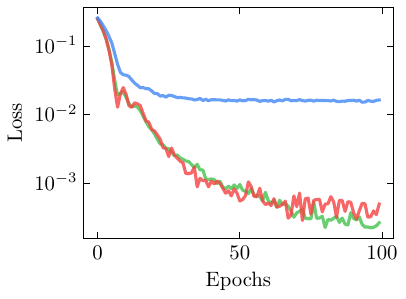}
  \caption{Predictions for the multidimensional function $f_{\rm ndim}$ with $N_{\rm dim}=4,6,8$
  from left to right. The exact predictions (black line)
  are compared with results obtained through noiseless simulation (green line), 
  noisy simulation (blue line), noisy simulation with mitigation applied to the 
  final predictions (yellow line) and real-time mitigated noisy simulation (red line).
  The effective depolarizing parameters $\lambda_{\text{eff}}$ are $0.22\pm0.02$ ($N_{\rm dim}=4$), 
  $0.31\pm0.03$ ($N_{\rm dim}=6$) and $0.41\pm0.02$ ($N_{\rm dim}=8$).
  The final predictions are computed averaging on $N_{\rm runs}=20$ predictions 
  calculated for each of the $N_{\rm data}=30$ points. The confidence intervals 
  are obtained using one standard deviation from the mean. The bottom plot shows 
  the loss function history for each training scenario.}
  \label{fig:cos_simulation}
  \end{figure*}

To better understand how the algorithm scales with the number of qubits
we study the problem of fitting a multi-dimensional function. In particular, we consider
\begin{equation}
f_{\rm ndim}(\bm{x}; \bm{\beta}) = \sum_{i=1}^{N_{\rm dim}} \bigl[ \cos{(\beta_i x_i)^{i}} + 
(-1)^{i-1} \beta_i x_i \bigr],
\label{eq:cosnd}
\end{equation}
where both data $\bm{x}$ and parameters $\bm{\beta}$ have dimension $N_{\rm dim}$
and the index $i$ runs over the problem dimensions. In particular, the model parameters
$\bm{\beta}$ are defined as equidistant point in the range $[0.5, 2.5]$, and they are 
kept fixed during the optimization. The target 
$f_{\rm ndim}$ is rescaled in order to occupy the range $[0,1]$. 
We consider $N_{\rm data}$ values for each $x_i \in \bm{x}$ homogeneously distributed in the range $[0, 1]$.
The ansatz is the $N_{\rm dim}$-qubit circuit of Fig.~\ref{fig:qpdf} with three layers.

As the dimensionality of the problem increases, and consequently, the number of qubits, the noise-induced bound is lower, 
hindering the description of the function in a region of its domain. 
By applying the RTQEM algorithm, we manage to achieve lower values of the loss function, thereby improving the quality 
of the fit (see Fig.~\ref{fig:cos_simulation}). These results are confirmed by computing the Mean Squared Error (MSE) metric,

\begin{equation}
\text{MSE} = \frac{1}{N_{\rm data}}\sum_{j = 1}^{N_{\rm data}}(\bar{y}_{\rm est}^j 
- y_{\rm meas}^j)^2 \ ,
\label{eq:MSE}
\end{equation}
where $\bar{y}_{\rm est}^j$ is the average estimate of $f(\bm{x}^j)$ over $N_{\rm runs}$.
The MSE associated to each fit is shown in Tab.~\ref{tab:sim_MSE}.

\begin{table}[h!]
\centering
\begin{tabular}{ccccc}
\hline \hline 
\textbf{Target} & $\text{MSE}_{\rm noiseless}$ & $\text{MSE}_{\rm noisy}$ & $\text{MSE}_{\rm fqem}$ & $\text{MSE}_{\rm rtqem}$\\
\hline
$u$ PDF    &  $0.008$  & $0.018$  & $0.023$ & $0.008$ \\
$\cos 4$d  &  $0.003$  & $0.043$  & $0.140$ & $0.003$ \\
$\cos 6$d  &  $0.002$  & $0.083$  & $0.214$ & $0.002$ \\
$\cos 8$d  &  $0.001$  & $0.118$  & $0.360$ & $0.004$ \\
\hline \hline
\end{tabular}
\caption{\label{tab:sim_MSE} Mean squared error distances between the target 
functions and the VQC fitting model trained under the different conditions of 
Tab.~\ref{tab:simulations}.}
\end{table}

Regarding the gradients, it is important to note that there are no significant differences between the raw gradients and the exact gradients (see Appendix~\ref{app:Gradients}). 
This means that we are in a regime where the loss concentration is not severe, and there is still room for error mitigation to improve trainability by mitigating 
other unwanted effects in the landscape due to the noise. 

\subsubsection{\label{sec:evolving_noise}Evolving-noise scenario}

To study the performance of the method with noise evolution, we consider a random change in the 
Pauli parameters of the noise model in each epoch. In particular, the initial 
parameters vector $\bm{q}^0=(q_{x}^0, q_{y}^0, q_{z}^0)$ is moved in its three-dimensional space following 
a procedure similar to a Random Walk (RW) on a lattice. Namely, each component 
$q_j$ is evolved from epoch $k$ to epoch $k+1$ as
\begin{equation}
q_j^{(k+1)} = q_j^{k} + r\delta,
\end{equation}
where $r\sim\{-1,+1\}$ and the step length is sampled from a normal distribution 
$\delta\sim\mathcal{N}(0,\sigma_{\delta})$. We refer to an evolution performing 
$N$ steps governed by $\sigma_{\delta}$ as $\text{RW}^{N}_{\sigma_{\delta}}$.
The readout noise parameter is kept fixed at the value of $q_{M}=0.005$.
In this evolving scenario, when the metric~\eqref{eq:metric_noise} exceeds
a certain threshold $\varepsilon_{\ell}$, the mitigation parameter $\lambda_{\text{eff}}$~\eqref{eq:lambda_eff} is updated. 

\begin{figure}[h]
  \centering
    \includegraphics[width=0.85\linewidth]{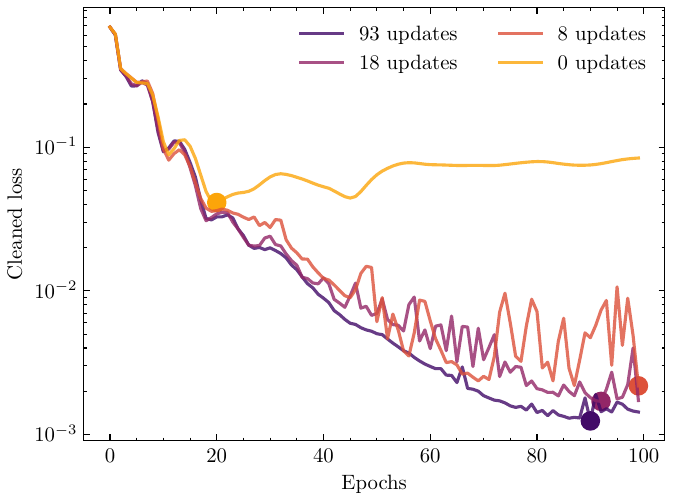}
    \caption{Four RTQEM training simulations sharing the same initial conditions. 
    The initial local Pauli parameters $\bm{q}^0 = (0.005, 0.005, 0.005)$ are evolved under 
    a $\text{RW}_{0.002}^{100}$. The readout noise parameter has been kept fixed to $q_M=0.005$, 
    four layers are used with $N_{\rm data}=30$ and $\eta=0.05$.
    For each training simulation a different noise threshold value was used: 
    $\varepsilon_{\ell}=\{0, 0.05, 0.1, 0.2\}$. As a result, 
    $\lambda_{\text{eff}}$ is re-learned $u=\{93, 18, 8, 0\}$ times, respectively. 
    We show the loss function values   
    computed using the training parameters at each iteration but deployed in a noiseless scenario.}
    \label{fig:noise_evolution}
\end{figure}

To evaluate the effect of relearning the noise on the training process, we track the evolution of the loss function of the $u$-quark PDF for various values of $\varepsilon_{\ell}$, as shown in Fig.~\ref{fig:noise_evolution}.
We aim for a reduction in the loss function to correspond to a closer approximation to the noise-free parameters. 
Therefore, we recalculate the loss function values at each iteration using the 
noisy training parameters, but in a noiseless simulation.
As the threshold decreases, the noise map is updated more frequently. 
It is expected that a lower threshold will enhance the training until it reaches a certain minimum value, characterized by the standard deviation of $\lambda_{\text{eff}}$.
Interestingly, even a few updates to the noise map can lead to significantly lower values for the loss function.
For instance, the difference between the minimum values of the loss function when updating the noise map 8 times as opposed to 93 times during a training of 100 epochs is $\mathcal{O}\left (10^{-3}\right )$.
This suggests that a minor additional classical computational cost can significantly improve the training. 

\subsection{\label{sec:hardware}Training on hardware}

We set up our full-stack gradient descent training on a superconducting
device hosted by the Quantum Research Center (QRC) in the Technology
Innovation Institute (TII). The high-level 
algorithm is implemented with \Qibo~\cite{Efthymiou_2021, Efthymiou_2022, Carrazza_2023,
stavros_efthymiou_2023_7736837} and then translated into pulses and executed on the
hardware through the \Qibolab~\cite{efthymiou2023qibolab, stavros_efthymiou_2023_7748527} framework (see Appendix~\ref{app:Native_gates}). 
The qubit calibration and characterization
have been performed using \Qibocal~\cite{pasquale2023opensource, andrea_pasquale_2023_7662185}.
In particular, we use one qubit of \texttt{Soprano}, a five-qubit chip constructed by QuantWare~\cite{quantware} and 
controlled using Qblox~\cite{qblox} instruments (see Appendix~\ref{app:lab_setup}). We refer to this device as \texttt{qw5q}.

The $u$-quark PDF for a fixed energy scale $Q_0$
is targeted using a four layer single-qubit circuit built following the ansatz presented above. 
We take $N_{\rm data}=15$ values of the 
momentum fraction $x$ sampled from the interval $[0,1]$.

\begin{figure}
  \centering
  \includegraphics[width=0.9\linewidth]{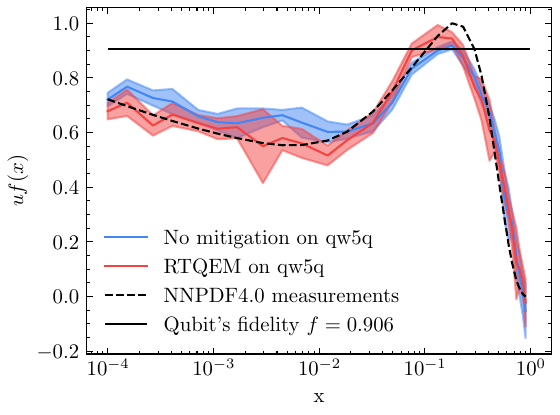}
  \includegraphics[width=0.9\linewidth]{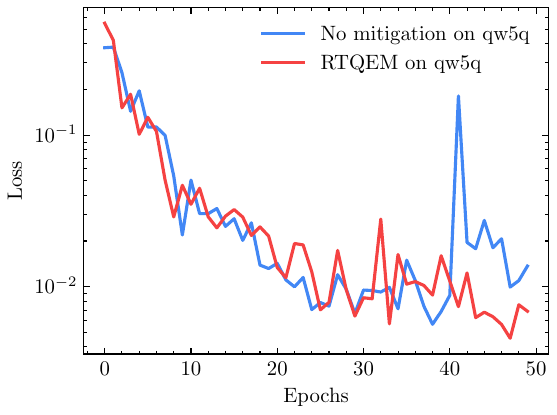}
  \caption{Above, estimates of $N_{\rm target}=30$ values of the $u$-quark PDF obtained 
  by training the best qubit of the \texttt{qw5q} chip. 
  The target values (black dashed line) are compared with our predictions after
  an unmitigated training (blue line) and a training with RTQEM (red line). The 
  estimations are computed averaging over $N_{\rm runs}=10$ predictions for each $x$
  with the trained $\bm{\theta}_{\rm best}$. The same prediction sets allow to 
  calculate the standard deviations of the estimates, which are then used to 
  draw the confidence intervals. Below, loss function history as function 
  of the optimization epochs. The effective depolarizing parameter is 
  $\lambda_{\rm eff} = 0.07 \pm 0.03$.
  }
  \label{fig:qw5q}
\end{figure}

An estimate to the bound imposed by the noise is provided by the assignment
fidelity of the used qubits, which are collected in dedicated runcards describing 
the current status of the QRC devices~\cite{qibolab_platforms_qrc}. 
As with the simulation, we adjust the function to span the range $[0,1]$.

\begin{center}
\begin{table}[ht]
\begin{tabular}{ccccccc}
\hline \hline
  \textbf{Param}          & $N_{\rm epochs}$ & $N_{\rm shots}$ & $N_{\rm train}$ & $N_{\rm params}$ & $\eta$ & \texttt{NumPy} seed\\
  \hline
  \textbf{Value}          & $50$             & $500$ & $15$ & $16$ & $0.1$ & $1234$\\  
\hline \hline
\end{tabular}
\caption{Hyper parameters of the gradient descent on \texttt{qw5q}}
\label{tab:sgd_qw5q}
\end{table}
\end{center}

We perform a gradient descent on the better calibrated qubit of \texttt{qw5q} using 
the parameters collected in Tab.~\ref{tab:sgd_qw5q}. The training has been performed for both the unmitigated and the RTQEM approaches. After training, we repeat 
$N_{\rm runs}=10$ times the predictions for each one of $N_{\rm target}=30$ target 
values of $x$ sampled logarithmically from $[0,1]$. The final estimate to the average prediction and its corresponding standard deviation are computed out of the $N_{\rm runs}$ repetitions.

The RTQEM process leads to better 
compatibility overall and, in particular, is able to overcome the bound set by the noise represented
as a black horizontal line, as shown in Fig.~\ref{fig:qw5q}.
Indeed, the mitigated fit leads to a smaller MSE compared to the unmitigated one,
as reported in Tab.~\ref{tab:hardware_mse}. 
This proves that the RTQEM procedure gives access to regions which are naturally forbidden 
to the raw training.
 
As a second test, we push forward the RTQEM training by performing 
a longer optimization. We use the same hyper-parameters of Tab.~\ref{tab:sgd_qw5q} 
but set $N_{\rm epochs}=100$, with the aim of finding 
more reliable parameters. We repeat the optimization twice, adopting the same initial 
conditions but changing the device. We use the aforementioned \texttt{qw5q} and 
a different five-qubit chip constructed by IQM~\cite{iqm} and controlled using Zurich~\cite{zurich} Instruments (see Appendix~\ref{app:lab_setup}).
We refer to this device as \texttt{iqm5q}. 

\begin{figure}
  \centering
  \includegraphics[width=0.9\linewidth]{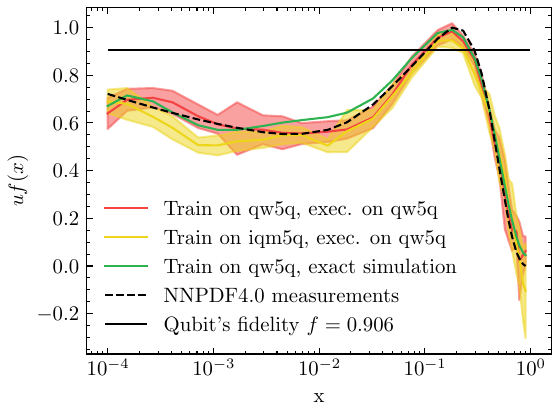}
  \caption{Estimates of $N_{\rm target}=30$ values of the $u$-quark PDF obtained 
  by training the better calibrated qubits of \texttt{qw5q} and  
  \texttt{iqm5q}, respectively with assignment fidelities $f_{\rm qw5q}=0.906$ 
  and $f_{\rm iqm5q}=0.967$. 
  The target values (black dashed line) are compared with our RTQEM predictions 
  obtained by training for  $N_{\rm epochs}=100$ on both \texttt{qw5q} 
  (red line) and \texttt{iqm5q} (yellow line). 
  We also show the predictions computed deploying in exact simulation mode the 
  best parameters obtained through RTQEM training on \texttt{qw5q} (green line).
  The final average and standard deviation of the predictions are computed out of $N_{\rm runs}=20$ repetition for each $x$
  using the parameters $\bm{\theta}_{\rm best}$ learned during training. 
  In particular, the $1\sigma$ confidence intervals are shown in the plot. The effective depolarizing parameter is 
  $\lambda_{\rm eff} = 0.08 \pm 0.02$.
  }
  \label{fig:benchs}
\end{figure}
\begin{center}
  \begin{table}[h]
  \begin{tabular}{ccccc}
  \hline \hline 
  \textbf{Training} & \textbf{Predictions} &  \textbf{Config.} & $N_{\rm epochs}$ & MSE \\
  \hline
  \texttt{qw5q} & \texttt{qw5q} & Noisy & $50$ & $0.0055$   \\     
  \texttt{qw5q} & \texttt{qw5q} & RTQEM & $50$ & $0.0042$   \\     
  \texttt{qw5q} & \texttt{qw5q} & RTQEM & $100$ & $0.0013$  \\     
  \texttt{iqm5q} & \texttt{qw5q} & RTQEM & $100$ & $0.0037$ \\   
  \texttt{qw5q} & \texttt{sim} & RTQEM & $100$ & $0.0016$ \\   
  \hline \hline
  \end{tabular}
  \caption{\label{tab:hardware_mse} MSE values for the models trained on the hardware. The column 
  ``Training'' indicates the device where the training took place, whereas the column 
  ``Predictions'' specifies the device where the model is deployed for testing.}
  \end{table}
\end{center}

If the parameters obtained through RTQEM procedure are noise independent, we expect them to be generally valid. Namely, 
the optimal parameters obtained for one device, should lead to a valid fit when 
deployed to a different one. 
This is illustrated in Fig.~\ref{fig:benchs}, where we report the results obtained 
by training individually on \texttt{qw5q} and on \texttt{iqm5q} with the same initial conditions, 
and then deploying the two sets of obtained parameters solely on \texttt{qw5q}. The plotted estimates are computed by averaging on $N_{\rm runs}=20$
repeated predictions. 

Finally, to further verify that the obtained parameters are indeed noise-independent, 
we deploy the model obtained by training on \texttt{qw5q} via RTQEM on an exact simulator
(green line in Fig.~\ref{fig:benchs}).

We calculate the MSE value for each described experiment following~\ref{eq:MSE}. 
All the results are collected in Tab.~\ref{tab:hardware_mse}, and confirm that the RTQEM training 
leads to noise-indipendent modelization.

\section{\label{sec:conclusion}Conclusion}

In this paper, we introduced a new Real-Time Quantum Error Mitigation (RTQEM) routine designed to enhance the training process of Variational Quantum Algorithms. 
We employed the Importance Clifford Sampling method at each learning step to mitigate noise in both the gradients of the loss function and the predictions. 
The RTQEM algorithm effectively reduces loss corruption without exacerbating loss concentration, thereby guiding the optimizer towards lower local minima of the loss function. 
We evaluated the RTQEM procedure using superconducting qubits and found that it improved the fit's consistency by surpassing the limitations imposed by the hardware's noise.

Our results demonstrate that the proposed algorithm effectively trains Variational Quantum Circuit (VQC) models in noisy environments. 
Specifically, if the system's noise remains constant or changes slowly, the noise map requires only a few updates during training, keeping the computational cost on par with the unmitigated training process. 

Notably, by mitigating noise during training, we can derive parameters that closely approximate those of a noise-free environment.
This adaptability allows us to deploy these parameters on a different device, even if it is subject to different noise.
This capability paves the way for the potential integration of federated learning with quantum processors.

The extension of this approach to other QML pipelines that use expected values as predictors, as well as to other QEM methods, presents an intriguing avenue for future research.
For instance, it can be applied to VQC models for supervised, unsupervised, and reinforcement learning scenarios in noisy environments.

\acknowledgments We would like to thank all members of the QRC lab for their support with the calibration of the devices.

We thank Juan Cereijo, Andrea Pasquale and Edoardo Pedicillo for the insightful discussions about the description of the devices used in this study.

This project is supported by CERN's Quantum Technology
Initiative (QTI). MR is supported by CERN doctoral program. AS acknowledges financial support through the Spanish Ministry of Science 
and Innovation grant SEV-2016-0597-19-4, the Spanish MINECO grant PID2021- 127726NB-I00, the Centro de Excelencia Severo Ochoa Program SEV-2016-0597 
and the CSIC Research Platform on Quantum Technologies PTI-001. AP was supported by an Australian Government Research Training
Program International Scholarship. SC thanks the TH
hospitality during the elaboration of this manuscript.

\bibliography{references}

\setcounter{equation}{0}\renewcommand\theequation{A\arabic{equation}}

\appendix
\onecolumngrid

\section{\label{app:Gradients}Gradients evolution}

\begin{figure*}[h]
  \centering
      \includegraphics[width=0.25\linewidth]{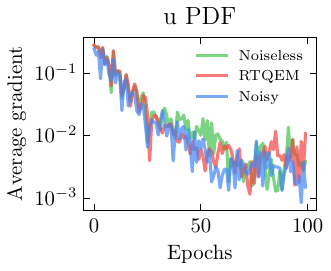}%
      \includegraphics[width=0.25\linewidth]{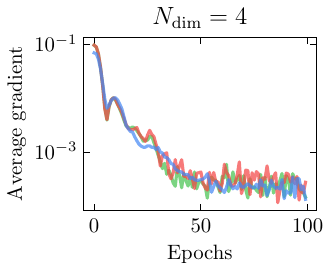}%
      \includegraphics[width=0.25\linewidth]{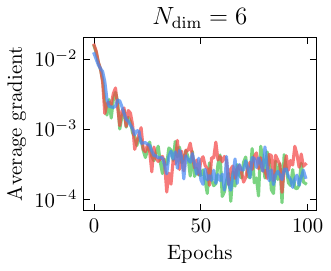}%
      \includegraphics[width=0.25\linewidth]{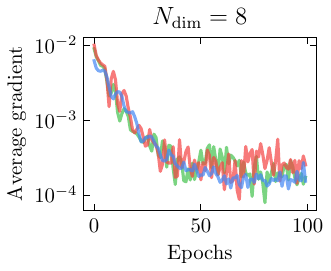}
  \caption{Average gradients as function of the optimization epochs. A noiseless simulation (green lines)
          is compared with unmitigated noisy simulation (blue lines) and RTQEM (red lines) 
          for $N_{\rm dim}=4,6,8$ from the left to the right plots.}
  \label{fig:cos_sim_gradients}
  \end{figure*}

During the VQC training, the noisy gradients are of the same magnitude as the exact ones, indicating that we are in a regime where exponential concentration is not severe, 
as shown in Fig.~\ref{fig:cos_sim_gradients}.

\section{\label{app:Native_gates}Native gates}

The native gates of the QRC superconducting quantum processors are $RX(\pm\pi/2)$, $RZ(\theta)$, and $CZ$ gates~\cite{qibolab_platforms_qrc}. 
They constitute a universal quantum gate set.
These gates are compiled into microwave pulses following a specific set of rules~\cite{efthymiou2023qibolab}.
For a circuit to be executable on hardware, it needs to be decomposed into these native gates. For instance, a general single-qubit unitary beaks into a sequence of five native gates,
\begin{equation}
  U(\theta,\phi,\lambda) = RZ(\phi)RX(-\pi/2)RZ(\theta)RX(\pi/2)RZ(\lambda) \ .
\end{equation} 

\section{\label{app:lab_setup}Qubits' parameters}

Relevant parameters of the qubits utilized in this study are presented in Tab.~\ref{tab:qubits}, including:

\begin{enumerate}[noitemsep]
\item the qubit transition frequency $f_{01} =\omega_q/2\pi$ from $\ket{0}$ to 
$\ket{1}$;
\item the bare resonator frequency $f_{\rm res}=\omega_{\rm res}/2\pi$;
\item the readout frequency $f_{\rm read}$ (coupled resonator frequency);
\item the energy relaxation time $T_1$;
\item the dephasing time $T_2$;
\item the time $\tau_g$ required to execute a single $RX$ gate;
\end{enumerate}

In the same table we also show the assignment fidelity $f= 1-[P(1|0) - P(0|1)]/2$, where $P(i|j)$ is a misclassification metric, counting the states prepared as 
$\ket{j}$ but measured as $\ket{i}$. This value is primarily due to the calibration status of the devices, rather than construction limitations.

\begin{center}
\begin{table}[ht]
\begin{tabular}{ccccccccc}
\hline \hline 
\textbf{Qubit} & $f_{01} ({\rm GHz})$ & $f_{\rm res} ({\rm GHz})$ & $f_{\rm read} ({\rm GHz})$ & $T_1 (\mu\text{s})$ & $T_2 (\mu\text{s})$ & $\tau_g ({\rm ns})$ & $f$ \\
\hline
\,\,qubit 4, \texttt{iqm5q} & $4.0978$ & $5.5047$ & $5.5150$ & $9.891$ & $3.700$ & $40$ & $0.967$ \\
qubit 3, \texttt{qw5q}  & $6.7599$ & $7.8000$ & $7.8028$ & $2.776$ & $1.139$ & $40$ & $0.906$\\
\hline \hline
\end{tabular}
\caption{\label{tab:qubits} Parameters of the \texttt{iqm5q} and \texttt{qw5q} qubits employed in this study.}
\end{table}
\end{center}

\end{document}